\documentclass[twocolumn,aps,floats,prl]{revtex4}
\usepackage{epsfig}

\begin{document}

\noindent
{\bf Castellano and Pastor-Satorras Reply:}
The Comment by Ha {\em et al.}~\cite{Ha06} criticizes our recent
result~\cite{Castellano06} that the contact process (CP) on uncorrelated
scale-free
(SF) networks does not behave according to heterogeneous mean-field
(MF) theory. The criticism is based on the following three claims:
(1) The relative density fluctuations discussed in Fig.~4
of~\cite{Castellano06} are well reproduced by a Gaussian ansatz, Fig.
1 of~\cite{Ha06}.
(2) A numerical estimate of the density decay for $\gamma=2.75$
agrees with the MF prediction $\theta=1/(\gamma-2)$.
(3) An estimate of the finite-size scaling (FSS)
exponent $\alpha=\beta/ \nu_\perp = 0.59(2)$ for the same $\gamma$ 
agrees with the MF conjecture $\alpha=1/(\gamma-1)$. 
We reply to these three points in the following paragraphs.

\begin{figure}[b]
  \epsfig{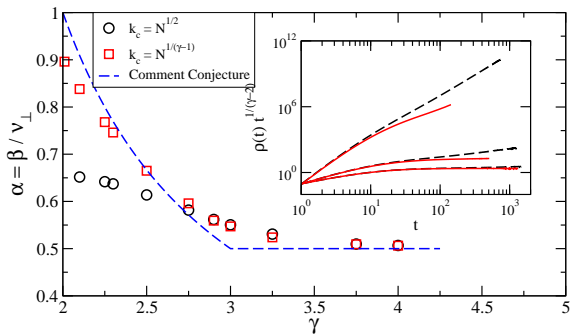}
  \caption{Exponent $\alpha=\beta/\nu_\perp$ 
%from FSS analysis at the
%critical point 
for RN SF networks for both cut-off scalings, compared with the
conjecture  $\alpha= 1/(\gamma-1)$.
Inset: Check of the scaling $\rho(t) \sim t^{-1/(\gamma-2)}$, for
$k_c \sim N^{1/2}$ (dashed lines) and   $k_c\sim N^{1/(\gamma-1)}$
(solid lines). Values of $\gamma$ (bottom to top): $2.75$, $2.50$ and $2.20$.}
%Simulation parameters as in Ref.~\protect\cite{Ha06}.}
  \label{Figure1}
\end{figure}

(1) The scaling form shown in Fig. 1 of~\cite{Ha06} seems
rather  interesting. However, it does not affect the main
conclusions of our work.
In fact, it proves that for almost all values of $k$, the
unscaled relative density fluctuations $r_k$ are larger than $1$ and the height
of the maximum diverges with the network size.  A diverging
ratio $r_k$ contradicts the MF assumption that the densities $\rho_k$ are
well defined quantities, and therefore naturally hints more towards  failure
of MF theory.
Concerning the numerical points
(2) and (3), we  note that the
critical point $p_c=0.4240$ quoted in~\cite{Ha06}  is not compatible
with our own estimate $p_c=0.4215(5)$ ~\cite{Castellano06}, hinting that a diffent network model is being used in~\cite{Ha06}.
%.  This leads us to conclude that the  authors of the
%Comment are using a network model different from
%ours, although no detail is provided at this respect.
%Neglecting for the moment this fact, 
In any case, the estimate of $\theta$
extracted from Fig.~2 in~\cite{Ha06} is utterly implausible:
the value is arbitrarily obtained from an extremely
short and noisy ``plateau'', whose existence appears more as
the effect of chance than a real physical feature.  
Concerning point (3), we are puzzled by the conjecture $\alpha=1/(\gamma-1)$,
of which no proof is given, except a mention to ``the hyperscaling
argument''~\cite{Ha06}.
%whose validity cannot be assessed since the authors provide
%no proof, except for a  reference to ``the hyperscaling
%argument''~\cite{Ha06}.  
We note, however, that it is possible to
obtain it from a trivial, although misleading, argument: in
Ref.~\cite{Castellano06} we observed that the MF
density of surviving particles at criticality should scale in uncorrelated SF
networks as
$\bar{\rho}_a \sim k_c^{-1}$, where $k_c$ is the degree cut-off. 
Assuming that $k_c\sim N^{1/2}$, corresponding to
\textit{real uncorrelated} networks \cite{mariancutofss},
our original result
$\alpha=1/2$ is recovered.
On the other hand, taking $k_c\sim N^{1/(\gamma-1)}$, which
corresponds to \textit{real  correlated} networks, yields the value
$\alpha=1/(\gamma-1)$. This result, however, makes little sense, since the MF
theory developed in our paper~\cite{Castellano06} (and presumably
in~\cite{Ha06}) deals with \textit{uncorrelated} networks.

In order to shed light on the true MF behavior of the model,
we have performed new simulations of the CP on the random
neighbors (RN) version of SF
networks, in which $p_c$ takes the exact MF value $1/2$ \cite{Castellano06}, considering 
the two different cut-off scalings discussed above.
%, namely $k_c \sim N^{1/2}$
%and $k_c \sim N^{1/(\gamma-1)}$. 
In the main plot of
Fig.~\ref{Figure1} we show results corresponding to the estimate of the
$\alpha$ exponent at $p_c$.
%We observe that both cut-off scalings provide almost
%identical results for $\gamma \ge 3$, while they deviate for
%smaller values. More importantly, 
We observe that the results for the realistic cut-off
$k_c \sim N^{1/2}$ deviate strongly from the conjecture $\alpha=1/(\gamma-1)$,
which is not even correct for the unphysical cut-off
(in uncorrelated networks) $k_c \sim N^{1/(\gamma-1)}$.
Finally, we consider a more natural way to
check the value of the $\theta$ exponent: since $\rho(t) \sim t^{-\theta}$,
the function $\rho(t) t^{\theta}$ should  show a plateau at
intermediate values of $t$ for the correct value of $\theta$.
In the inset of Fig.~\ref{Figure1} the MF value $\theta=1/(\gamma-2)$
is tested, for both cut-offs considered.
As we can see, the MF value is reasonable for large $\gamma$, but it
fails completely for $\gamma$ close to $2$. 

From this  numerical evidence we conclude that the MF conjecture presented
in the Comment is at best only approximately valid for the unphysical case
of uncorrelated networks with cut-off $k_c \sim N^{1/(\gamma-1)}$,
which can only be constructed in the RN version of SF networks, while the
MF prediction for $\theta$ fails completely for small $\gamma$ and any $k_c$.
Therefore, the main conclusion of paper~\cite{Castellano06}, the
invalidity of MF theory for \textit{real  uncorrelated} SF networks,
remains unchallenged.

\vspace{1eX}

\noindent
Claudio Castellano$^1$ and Romualdo Pastor-Satorras$^2$\\
\indent $^1$CNR-INFM, SMC, Dipartimento di Fisica, \\
\indent Universit\`a di Roma ``La Sapienza'' \\
\indent P.le Aldo Moro 2, I-00185 Roma, Italy \\
\indent $^2$Departament de F\'\i sica i Enginyeria Nuclear \\
\indent Universitat Polit\`ecnica de Catalunya, \\
\indent Campus Nord B4, 08034 Barcelona, Spain

%\vspace{-0.9cm}

\end{document}